\documentclass[aps,pra,superscriptaddress,twocolumn]{revtex4-2}
\usepackage{bbm}
\usepackage{mathrsfs}
\usepackage{amsmath}
\usepackage{amsfonts}
\usepackage[colorlinks=true,citecolor=blue,anchorcolor=blue]{hyperref}
\usepackage{graphicx,epstopdf}
\usepackage{subfigure}
\usepackage{epsfig}
\usepackage{dcolumn}
\usepackage{bm}
\usepackage{color}
\usepackage{natbib}
\usepackage{amssymb}
\usepackage{xcolor}
\usepackage{braket}
\usepackage[T1]{fontenc}
\usepackage[utf8]{inputenc}
\usepackage{braket}
\usepackage{physics}
\usepackage{amsmath, amssymb}

\usepackage[capitalise]{cleveref}
\crefname{section}{Sec.}{Secs.}
\crefname{appendix}{App.}{Apps.}

\begin{document}
	\title{Remote state preparation of a single-spin state via hybrid spin-photon entanglement}
	\author{Yunge Jiang}
\thanks{These authors contributed equally to the work}
\affiliation{Department of Physics, Wenzhou University, Zhejiang 325035, China}

\author{Yehan Yu}
\thanks{These authors contributed equally to the work}
\affiliation{Department of Physics, Wenzhou University, Zhejiang 325035, China}

\author{Lili Song}
\affiliation{Department of Physics, Wenzhou University, Zhejiang 325035, China}

\author{Yan Mu}
\affiliation{Department of Physics, Wenzhou University, Zhejiang 325035, China}

\author{Chaoyun Peng}
\altaffiliation{chaoyun@wzu.edu.cn}
\affiliation{Information Technology Center, Wenzhou University, Zhejiang 325035, China}

\author{Mingfeng Wang}
\altaffiliation{mfwang@wzu.edu.cn}
\affiliation{Department of Physics, Wenzhou University, Zhejiang 325035, China}

\date{\today }

	\begin{abstract}
Remote state preparation is a fundamental quantum-information protocol that exploits prior knowledge of a target state to reduce the communication resources required for quantum-state distribution. Here, we propose a remote-state-preparation protocol for a stationary electron-spin qubit using hybrid entanglement between the electron spin and a coherent-state light pulse, generated by spin-dependent reflection from a spin–cavity system. Unlike single-photon encodings, the coherent-state encoding is naturally tolerant of photon loss: a loss channel attenuates the coherent-state amplitude, leading to a gradual reduction in preparation fidelity rather than probabilistic protocol failure. The central difficulty---implementing the projection onto superpositions of nonorthogonal coherent states required in the conventional protocol---is circumvented by coupling the transmitted pulse to an auxiliary spin, followed by homodyne detection and a spin projective measurement. The protocol therefore avoids direct coherent-state-superposition measurements and photon-number-resolving detection. In the ideal limit, deterministic preparation of a particular class of target states can be achieved using only one bit of classical communication. We analytically quantify the effects of coherent-state nonorthogonality, fiber loss, and spin dephasing on the preparation fidelity. Increasing the coherent-state amplitude improves state distinguishability and hence the preparation fidelity, but also enhances which-branch information leakage caused by photon loss, resulting in an optimal amplitude at each transmission distance. For experimentally relevant parameters, the optimized average fidelity remains well above the classical benchmark over distances of tens of kilometers. We also discuss a possible implementation using diamond nitrogen-vacancy centers.

	\end{abstract}
	\maketitle
	
\section{Introduction}

	Quantum state transfer~\cite{doi:10.1126/science.1103346,Northup2014,christandl2004perfect} is one of the central themes in quantum information science~\cite{vedral2006introduction,nielsen2010quantum} and serves as a key foundation for quantum communication~\cite{gisin2007quantum,BENNETT20147}, distributed quantum computing~\cite{caleffi2024distributed,barral2025review,PhysRevA.76.062323,monroe2013scaling}, and quantum networks~\cite{PhysRevA.80.022339}. Efficiently transferring quantum states between spatially separated quantum nodes~\cite{PhysRevLett.78.3221,Kimble2008} is crucial for remote quantum information processing and the sharing of quantum resources. Among quantum information transmission protocols~\cite{gisin2007quantum,RevModPhys.77.513}, remote state preparation (RSP) ~\cite{PhysRevLett.87.077902,PhysRevA.62.012313} has attracted considerable attention because the sender’s prior knowledge of the target state enables a more resource-efficient implementation. In particular, RSP can require less classical communication than the teleportation of an unknown quantum state~\cite{PhysRevLett.70.1895}.
	
	Since Lo's seminal proposal of single-qubit RSP~\cite{PhysRevA.62.012313},
RSP has been extensively studied with various entanglement resources
and in diverse quantum systems. Previous studies have shown that RSP can reduce the classical communication cost for certain classes of target states~\cite{pati2000minimum,PhysRevLett.87.077902,hua2016deterministic,w20080405}. Subsequently, studies have been gradually extended to arbitrary pure states~\cite{fannes1994finitely,PhysRevA.65.012310,SCHUMAKER1986317}, mixed states~\cite{xiang2005remote,verstraete2001maximally,liang2019quantum}, multipartite states~\cite{wallnofer2019multipartite,PhysRevA.88.042307}, and high-dimensional states~\cite{mclaren2015realising,kues2017chip}. Various RSP variants
were also developed, including deterministic~\cite{an2011deterministic,li2016deterministic},
probabilistic~\cite{wei2014two,ping2008remote}, controlled~\cite{chen2012controlled,liu2014controlled}, and joint
RSP~\cite{nguyen2008joint,xiao2011joint}. More recently, further advances have been made in RSP using non-maximally entangled channels~\cite{xin2023deterministic,yi2002remote}, enhancing robustness against noise~\cite{zhou2021effect}, and demonstrating proof-of-principle implementations~\cite{liu2007experimental}.
	
Advances in experimental quantum information technologies~\cite{zhang2024tutorial,wehner2018quantum} have enabled RSP to be investigated and demonstrated across a variety of physical platforms, including photonic systems~\cite{liu2007experimental}, nuclear magnetic resonance (NMR) systems~\cite{peng2003experimental}, trapped ions~\cite{rowe2002transport}, superconducting quantum circuits~\cite{PhysRevX.6.031036}, and solid-state spin systems~\cite{kurokawa2022remote}. These platforms offer complementary advantages in coherence time, precise control, long-distance transmission, and integrability, thereby providing a versatile foundation for practical RSP implementations~\cite{wehner2018quantum}. In particular, photonic systems are well suited for transmitting quantum information over long distances~\cite{duan2001long}, whereas solid-state spin systems are promising candidates for quantum-network nodes because of their superior local quantum-storage and information-processing capabilities~\cite{doherty2013nitrogen,awschalom2018quantum}. A natural architecture for long-distance quantum communication therefore employs stationary spin qubits as network nodes and spin-photon hybrid entanglement as a quantum resource for quantum-state transfer and RSP~\cite{PhysRevLett.81.5932,Kimble2008}. Along this direction, Rosenfeld \textit{et al.} experimentally demonstrated the remote preparation of a single-spin state using the hybrid entanglement between an atomic spin and a single photon~\cite{PhysRevLett.98.050504}. However, because the protocol requires transmission of the entangled photon through a communication channel, photon loss can lead to protocol failure and thereby limit the achievable distance of RSP. This limitation motivates the development of alternative RSP schemes that are \emph{less sensitive to photon loss} and thereby extend the achievable distance.
	\begin{figure*}[t]
		\centering
		\includegraphics[width=0.92\textwidth]{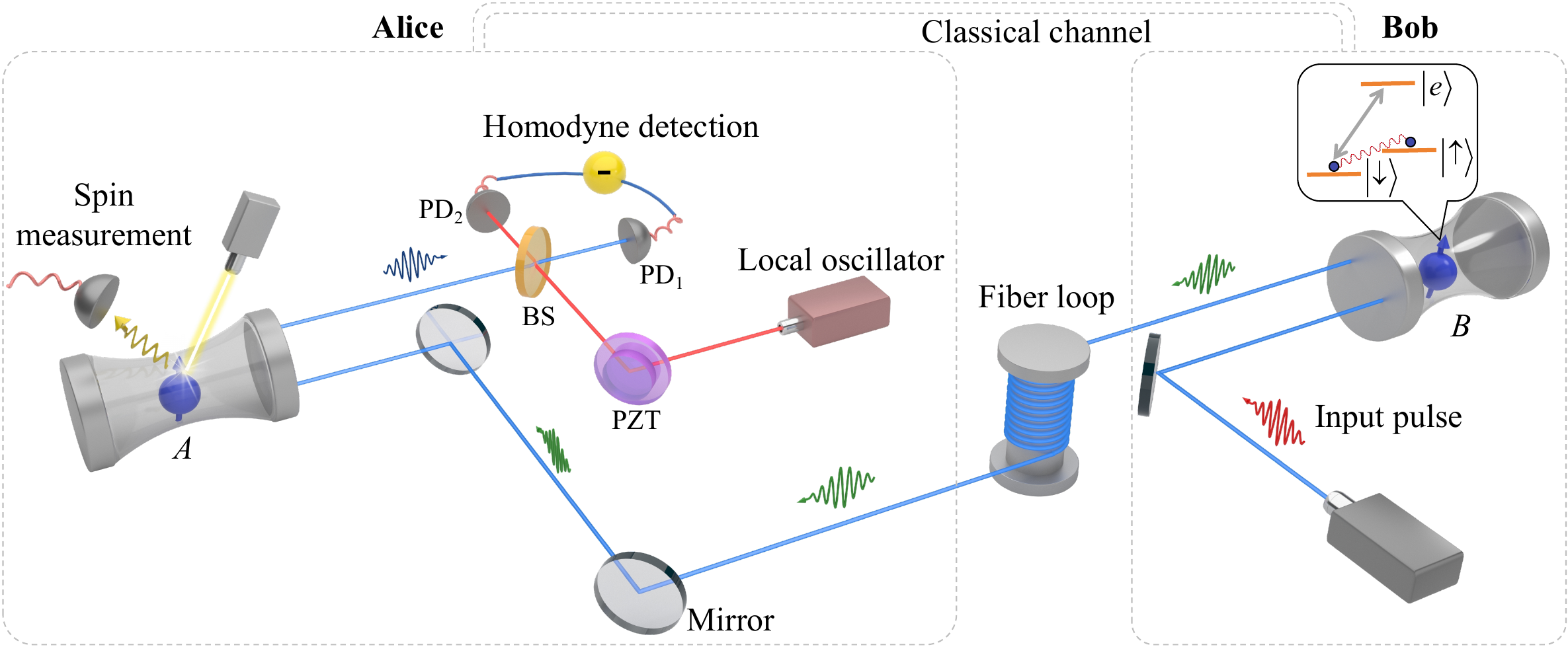}
		\caption{Schematic illustration of spin RSP based on hybrid spin-photon entanglement. A single $\Lambda$-type spin $B$ is confined in a high-finesse optical cavity at Bob's site, which is resonant with the $\ket{\downarrow}\leftrightarrow|e\rangle$ transition and far detuned from the $\ket{\uparrow}\leftrightarrow|e\rangle$ transition. An optical pulse initially prepared in a coherent state is reflected from the spin-cavity system, resulting in a hybrid spin-photon entangled state. The optical pulse carrying the quantum correlations is transmitted through an optical fiber to Alice---the receiving party, where it enters a second cavity with an auxiliary electron spin $A$ and interacts via linear cavity quantum electrodynamics. Projective measurements are performed on the optical field in the position basis and on spin $A$ in a customized orthogonal basis matched to the target state. With one-bit feedback via the classical channel, the remote spin $B$ at Bob’s site is deterministically prepared for target states within the accessible set.}
		\label{fig:rsp_hybrid}
	\end{figure*}

In this work, we propose a scheme for the remote preparation of a single-spin state that is robust against transmission loss. Unlike most previous RSP protocols based either on discrete-variable (DV)~\cite{PhysRevLett.87.077902,PhysRevX.6.031036} or continuous-variable (CV)~\cite{PhysRevA.72.052315,han2022remote,PhysRevLett.82.1784} systems, our scheme employs a hybrid system consisting of a DV spin and a CV optical field. Such hybrid light-spin entanglement is generated simply by reflecting a coherent-state pulse from a cavity containing a single spin. In contrast to single-photon-based protocols, where photon loss prevents successful state preparation, the multiphoton nature of the coherent-state pulse allows our protocol to tolerate partial photon loss, albeit with reduced fidelity of state preparation. Consequently, the protocol requires no heralding of successful transmission and can, in principle, operate deterministically. We also propose a new approach based on an auxiliary atom-cavity system that enables arbitrary projective measurements on the CV optical state using only homodyne detection of the optical field and a projective measurement of the auxiliary spin, thereby obviating the need for nonlinear optical elements~\cite{butcher1990elements,duan2001long,PhysRevLett.93.250502,PhysRevLett.101.253601} and enhancing the feasibility of the scheme. Our approach provides a viable route to remote state preparation and long-distance quantum state transfer in distributed hybrid quantum networks.

The rest of the paper is organized as follows. In~\cref{sec2}, we describe our hybrid spin‑photon platform and derive the spin‑dependent reflection dynamics in a spin-cavity system for generating hybrid entanglement between a stationary spin qubit and propagating coherent‑state pulses. In~\cref{sec3}, we first present our remote state preparation protocol, and then
 the average preparation fidelity is analyzed by taking the non‑orthogonality of coherent states into consideration. In~\cref{sec4}, we investigate the dominant decoherence channels including fiber photon loss and spin dephasing, revealing the competition between coherent‑state distinguishability and channel dissipation. We also discuss feasible experimental realization with diamond NV centers. Finally, a summary is provided in~\cref{sec5}. The Appendix contains full derivations for the concurrence of conditional two‑spin entangled state and the average fidelity over all target single‑spin states.

\section{Spin-light interaction model\label{sec2}}
We consider a hybrid quantum system consisting of a single three-level spin coupled to a high-finesse optical cavity and a propagating optical pulse prepared in a coherent state. This system provides the physical platform for generating the hybrid spin-photon entanglement required for our RSP protocol. The spin serves as the stationary target qubit, whereas the optical pulse acts as a flying quantum information carrier that distributes quantum correlations over long distances.
	
As shown in Fig.~\ref{fig:rsp_hybrid}, the electron spin possesses two stable ground states, denoted by	$\ket{\uparrow}$ and $\ket{\downarrow}$, which encode the logical qubit. The cavity mode resonantly couples the $\ket{\downarrow}$ state to the excited state $\ket{e}$. To create hybrid quantum entanglement, a coherent-state pulse $\ket{\alpha}$
is directed onto the cavity. The resulting spin-dependent reflection imparts a conditional phase shift to the optical field. For a narrow-bandwidth input pulse that is nearly resonant with the cavity, we obtain~\cite{PhysRevA.111.012438,PhysRevA.79.032303,PhysRevA.83.054305,PhysRevA.72.032333}
			\begin{eqnarray}
			\ket{\uparrow}\ket{\alpha} \xrightarrow{\text{reflection}} \ket{\uparrow}\ket{\alpha},
			\ket{\downarrow}\ket{\alpha} \xrightarrow{\text{reflection}} \ket{\downarrow}\ket{-\alpha}.\label{eq:spin_phase}
		\end{eqnarray}
This equation captures the essential physics of the spin-dependent phase shift: when the spin is in the state $\ket{\uparrow}$, the optical pulse is reflected without a phase shift, whereas when the spin is in the state $\ket{\downarrow}$, the pulse acquires a $\pi$ phase shift, resulting in the transformation $\ket{\alpha}\to\ket{-\alpha}$. A detailed derivation of this result based on the input-output formalism of cavity QED is given in Ref.~\cite{PhysRevA.111.012438}. The transformations (\ref{eq:spin_phase}) serve as the starting point for implementing the current scheme.
	
	To generate the entangled state used for RSP, we first initialize the electron spin in an equal superposition state:
	\begin{equation}
		\ket{\psi}_{S} = \frac{1}{\sqrt{2}} \left(\ket{\uparrow} + \ket{\downarrow}\right ).
	\end{equation}
Next, an optical pulse prepared in the coherent state $|\alpha\rangle$ is sent to interact with the spin-cavity system. According to the spin-dependent transformation in Eq.~\eqref{eq:spin_phase}, the joint spin-photon state after reflection becomes the following entangled state:
	\begin{equation}\label{eq:hybrid_ent}
		\ket{\Psi}_{E} = \frac{1}{\sqrt{2}} (\ket{\uparrow}\ket{\alpha} + \ket{\downarrow}\ket{-\alpha}).
	\end{equation}
This state represents a unique form of hybrid entanglement: a DV spin qubit is entangled with a CV photonic coherent state, which serves as the quantum channel for the subsequent RSP protocol. A key feature of the coherent-state basis is its intrinsic non-orthogonality. The overlap between the two logical states $\ket{\alpha}$ and $\ket{-\alpha}$ is given by
\begin{equation}
\bra{\alpha}\ket{-\alpha} = e^{-2\abs{\alpha}^2}.
\end{equation}
For sufficiently large coherent-state amplitudes, their overlap,
$\bra{\alpha}\ket{-\alpha} = e^{-2\abs{\alpha}^2}$, becomes exponentially small. In particular, for $|\alpha| \geq 2$, one has
$|\langle \alpha | {-}\alpha \rangle| \leq e^{-8} \approx 3.35 \times 10^{-4}$,
so that  $\ket{\alpha}$ and $\ket{-\alpha}$ can be regarded as nearly orthogonal. In this regime, the hybrid entangled state in Eq.~\eqref{eq:hybrid_ent} can be effectively treated as a standard DV Bell state, with $\ket{\alpha}$ and $\ket{-\alpha}$ forming an effective qubit basis for the photonic mode. From the perspective of quantum-information encoding, the nonorthogonality of coherent states may be disadvantageous. However, for quantum-state transmission, coherent-state encoding offers a distinct advantage over DV encodings. For example, in single-photon polarization encoding~\cite{PhysRevLett.92.127902}, the probability that the photon survives a lossy channel decreases exponentially with the transmission distance. By contrast, a pure-loss channel with transmissivity $\eta$ maps
$\ket{\alpha}$ onto $|\sqrt{\eta}\alpha\rangle$. Thus, channel loss  only attenuates the coherent-state amplitude, while the output remains a coherent state rather than being probabilistically erased. As we show below, this property is crucial for realizing remote quantum-state preparation over long distances.

\section{Remote quantum state preparation\label{sec3}}
\label{sec:rsp}
\subsection{Principle of RSP}
We now introduce the protocol for remotely preparing a prescribed
quantum state at a distant node. The protocol involves two spatially separated parties, Alice and Bob. Bob possesses a spin qubit, labeled $B$. The goal is for Alice to remotely prepare a target state on Bob's spin qubit using shared entanglement, local operations, and classical communication~\cite{PhysRevLett.70.1895}. The target state to be prepared is of the form \begin{equation}
    \ket{\psi_T}_{B}
    =  a\ket{\uparrow}_B
    + b\ket{\downarrow}_B,
    \qquad
    |a|^{2}+|b|^{2}=1,
    \label{eq:target_state}
\end{equation}
where the complex amplitudes $a$ and $b$ are known to Alice but unknown to Bob. As shown in Fig.~\ref{fig:rsp_hybrid}, to establish the quantum channel, Bob first prepares spin $B$ in an equal superposition state and then directs the optical pulse described above onto the cavity, allowing it to interact with the spin-cavity system. The reflected pulse is subsequently transmitted to Alice through an optical fiber, thereby establishing spin-photon entanglement shared between Alice and Bob. The resulting joint state is given by
\begin{equation}
\ket{\Psi}_{AB}=\frac{1}{\sqrt{2}}\bigl(\ket{\uparrow}_B\ket{\alpha}_A+\ket{\downarrow}_B\ket{-\alpha}_A\bigr).\label{eq6}
\end{equation}

Before presenting our protocol, we briefly review the conventional approach to remote state preparation based on hybrid spin-photon entanglement in Eq.~(\ref{eq6}). Using her classical knowledge of the target-state amplitudes $a$ and $b$, Alice performs a corresponding local operation and subsequently measures her subsystem in an appropriate basis. For the target state defined in Eq.~(\ref{eq:target_state}), we introduce the following basis:
\begin{align}
\ket{\mu_{0}}_A=a^{*}\ket{\alpha}_A+b^{*}\ket{-\alpha}_A,
\ket{\mu_{1}}_A = b\ket{\alpha}_A-a\ket{-\alpha}_A.
\label{eq7}
\end{align}
For sufficiently large $|\alpha|$, the overlap between the two states is exponentially suppressed, allowing them to be treated as approximately orthogonal. Thus, the state in Eq. (\ref{eq6}) can be expressed in terms of the approximately orthogonal basis in Eq. (\ref{eq7}) as
\begin{equation}
\ket{\Psi}_{AB}
=
\frac{1}{\sqrt{2}}
\bigl(
\ket{\mu_{0}}_A
\ket{\psi_T}_{B}
+
\ket{\mu_{1}}_A
\ket{\psi_{T_\perp}}_B
\bigr),
\label{eq8}
\end{equation}
where
\begin{equation}
\ket{\psi_{T_\perp}}_B
=
b^{*}\ket{\uparrow}_B - a^{*}\ket{\downarrow}_B.
\label{eq9}
\end{equation}
Equation~(\ref{eq8}) establishes an explicit correspondence between Alice's measurement outcomes and the resulting conditional states at Bob's node.
If Alice obtains the outcome corresponding to $\ket{\mu_{0}}_A$, Bob's subsystem is projected directly onto the target state $\ket{\psi_T}_{B}$. Alice then communicates the measurement outcome to Bob via a classical channel, and no further operation is required on Bob's subsystem.
The probability of this outcome is $\bigl\lVert _A\bra{\mu_0} \ket{\Psi}_{AB} \bigr\rVert^2 =1/2$.
If Alice obtains the outcome corresponding to $\ket{\mu_{1}}_A$, Bob's subsystem is instead projected onto the conditional state given in Eq.~(\ref{eq9}). For a general target state with arbitrary complex amplitudes, the corresponding conditional state cannot, in general, be transformed into $\ket{\psi_T}_{B}$ by a fixed unitary operation independent of the target-state parameters~\cite{pati2000minimum}. Consequently, the remote preparation of an arbitrary target state can only be achieved probabilistically. However, for a restricted ensemble of target states---for example, states with real amplitudes $a$ and $b$ --- the protocol can be deterministic, because applying the fixed, target-independent Pauli-$Y$ operator [that is, $\sigma_y=i(\ket{\downarrow}_B\bra{\uparrow}-\ket{\uparrow}_B\bra{\downarrow})$] to $\lvert\psi_{T_\perp}\rangle_B$ maps it onto the target state, up to a physically irrelevant global phase. In this case, Alice only needs to send one classical bit to Bob, informing him whether the Pauli-$Y$ correction should be applied. Therefore, the entire RSP protocol consumes one cbit of classical communication and one ebit of shared entanglement.

\subsection{Physical realization of the projection measurement}
As is evident from the protocol described above, in addition to the generation and
 distribution of the quantum channel, a central experimental challenge in
hybrid-entanglement-based RSP is the implementation
of the required projective measurements on the optical subsystem. Since the optical qubit is encoded in the nonorthogonal coherent states
$\ket{\alpha}$ and $\ket{-\alpha}$ of a continuous-variable
optical mode, implementing projective measurements onto arbitrary
superpositions within this encoded subspace---for example, the even and odd
coherent-state superpositions
$\mathcal{N}_{\pm}\bigl(\ket{\alpha} \pm \ket{-\alpha}\bigr)$---remains
experimentally challenging. Such measurements generally require dedicated
measurement resources, such as photon-number-resolving or parity detection,
adaptive photodetection, and effective nonlinear interactions~\cite{Cook2007,Jeong2001,Ralph2003,Ourjoumtsev2006}.
Here, we propose an alternative implementation of the projective
measurements required by the hybrid RSP protocol using an auxiliary
cavity-atom system. As shown below, our scheme requires only homodyne
detection (HD) of the optical mode and a single-spin measurement, both of which
rely on well-established experimental techniques, thereby substantially
simplifying the experimental implementation.

Our projective measurement scheme is illustrated in Fig.~\ref{fig:rsp_hybrid}.
After being received by Alice, the optical pulse is directed into a second cavity,
which contains atom $A$ that is initially prepared in the equal superposition state $\frac{1}{\sqrt{2}}(\ket{\uparrow}_A+\ket{\downarrow}_A)$.  Because the cavity--atom system is identical to the
first one, the pulse undergoes the same evolution, resulting in the following
tripartite state:
\begin{eqnarray}
\ket{\Psi} &=& \frac{1}{2}\Bigl[\ket{\uparrow}_A\bigl(\ket{\alpha}_A\ket{\uparrow}_B + \ket{-\alpha}_A\ket{\downarrow}_B\bigr) \nonumber\\
&&+ \ket{\downarrow}_A\bigl(\ket{-\alpha}_A\ket{\uparrow}_B + \ket{\alpha}_A\ket{\downarrow}_B\bigr) \Bigr].\label{eq10}
\end{eqnarray}
After the cavity-light interaction, the optical mode is
subjected to HD of its $x$ quadrature. Assuming that $\alpha$
is real, the coherent states can be expanded in the quadrature basis as
\begin{equation}
\ket{\pm\alpha}_A
=
\int dx\,\psi_{\pm}(x)\ket{x}_A,
\end{equation}
where
\begin{equation}
\psi_{\pm}(x)
=
\frac{1}{\pi^{1/4}}
\exp\left[
-\frac{1}{2}
\left(x\mp\sqrt{2}\alpha\right)^2
\right].
\end{equation}
Substituting these expansions into Eq.~(\ref{eq10}) gives
\begin{equation}
\begin{split}
\ket{\Psi}
=
&\frac{1}{2}\int dx\,
\Bigl [
\psi_{+}(x)
\left(
\ket{\uparrow}_A\ket{\uparrow}_B
+
\ket{\downarrow}_A\ket{\downarrow}_B
\right)
\\
&+
\psi_{-}(x)
\left(
\ket{\uparrow}_A\ket{\downarrow}_B
+
\ket{\downarrow}_A\ket{\uparrow}_B
\right)
\Bigr]\ket{x}_A.
\end{split}
\end{equation}
Therefore, conditioned on a homodyne outcome $x_m$, the two atoms are projected
onto the normalized state
\begin{align}
\ket{\Psi(x_m)}_{AB}
=
\frac{\psi_{+}(x_m)\ket{\Phi^{+}}_{AB} + \psi_{-}(x_m)\ket{\Psi^{+}}_{AB}}
{\sqrt{\bigl|\psi_{+}(x_m)\bigr|^2 + \bigl|\psi_{-}(x_m)\bigr|^2}},
\label{eq14}
\end{align}
where we have defined the atomic Bell states
\begin{eqnarray}
\ket{\Phi^{+}}_{AB}
&=&
\frac{1}{\sqrt{2}}
\left(
\ket{\uparrow}_A\ket{\uparrow}_B
+
\ket{\downarrow}_A\ket{\downarrow}_B
\right),\\
\ket{\Psi^{+}}_{AB}
&=&
\frac{1}{\sqrt{2}}
\left(
\ket{\uparrow}_A\ket{\downarrow}_B
+
\ket{\downarrow}_A\ket{\uparrow}_B
\right).
\end{eqnarray}

Apparently, the state given in Eq.~(\ref{eq14}) is an entangled state. To
quantify the degree of entanglement, we calculate the concurrence of the
spin system~\cite{Hill1997,Wootters1998}, obtaining (see Appendix A for details)
\begin{eqnarray}
C\left( {x_m,\alpha } \right) = \left| {\tanh \left( {2\sqrt 2 \alpha x_m} \right)} \right|.
\label{eq17}
\end{eqnarray}
The concurrence is plotted in Fig.~\ref{fig2} as a function of the measurement outcome $x_m$. As can be seen from the figure, a large degree of
entanglement can be generated when $|x_m|$ is sufficiently large. For a relatively large value of
$\alpha$, such as $\alpha=2$, the probability of obtaining a maximally
entangled state after the measurement approaches unity. Specifically,
depending on the measurement outcome, the post-measurement state of the atomic
system collapses onto
\begin{eqnarray}
\left|\Psi\right\rangle_{AB}
\simeq
\begin{cases}
\ket{\Phi^{+}}_{AB}, & x_m \gg 0,\\
\ket{\Psi^{+}}_{AB}, & x_m \ll 0.
\end{cases}
\end{eqnarray}
By performing HD on the output optical field, we successfully
convert the atom-light entanglement into entanglement between the two
atoms. In particular, if the resulting atomic state is
$\ket{\Psi^{+}}_{AB}$, Alice can then apply a Pauli-$X$ gate to her
qubit, thereby transforming the state according to
$\ket{\Psi^{+}}_{AB}
\longrightarrow
\ket{\Phi^{+}}_{AB}.$
Consequently, irrespective of the outcome of the optical-field
measurement, the two atoms can be deterministically prepared in the maximally entangled
state $\ket{\Phi^{+}}_{AB}$, up to a local operation on Alice's qubit.

Using the spin-spin entanglement $\ket{\Phi^{+}}_{AB}$, Alice subsequently performs a projective measurement on atom $A$ in the
basis $\{\ket{{\mu}_{S0}}_A\equiv a^{*}\ket{\uparrow}_A+ b^{*}\ket{\downarrow}_A, \ket{{\mu}_{S1}}_A\equiv b\ket{\uparrow}_A- a\ket{\downarrow}_A\}$.
Such measurement can be realized by applying an appropriate driving field
to atom $A$, followed by a measurement in the
$\{\ket{\uparrow}_A,\ket{\downarrow}_A\}$ basis, i.e., a $\sigma_z$
measurement in the rotated basis. Alice then communicates the
measurement outcome to Bob through a classical channel. For real‑valued coefficients $a,b$, based on the measurement outcome, Bob applies the corresponding unitary operation to his qubit, either the identity operation
$I$ or a rotation about the $y$ axis, thereby recovering the target state $\ket{\psi_T}_B$ and completing the RSP.

\subsection{RSP Fidelity}
In the above analysis, we assumed that the coherent states
$\ket{\alpha}$ and $\ket{-\alpha}$ are perfectly distinguishable.
For finite $\alpha$, however, they are nonorthogonal, with overlap $
\bra{\alpha}\ket{-\alpha} = e^{-2\abs{\alpha}^2}$.
This nonzero overlap limits the distinguishability of the two states
and consequently affects the performance of the RSP protocol. We
quantify this effect by evaluating the RSP fidelity,
\begin{equation}
\mathcal{F}
=
{}_B\!\bra{\psi_T}
\rho_B^{\mathrm{out}}
\ket{\psi_T}_B.
\end{equation}
where $\rho_B^{\mathrm{out}}$ is Bob's output state after the conditional
operation. Starting from the nonmaximally entangled state given in Eq. (\ref{eq14}), we implement the RSP protocol and derive the fidelity as (see Appendix B)
\begin{equation}
\mathcal{F}=\frac{\bigl[1+2\text{Re}(ab^{*}) e^{-2\sqrt{2}\alpha x_m}\bigr]^2}{1+e^{-4\sqrt{2}\alpha x_m}+4\text{Re}(ab^{*}) e^{-2\sqrt{2}\alpha x_m}}.
\label{eq:f_xm}
\end{equation}
This expression quantifies the reduction in the RSP fidelity arising from the nonmaximal entanglement of the quantum channel. The fidelity is determined by the distinguishability parameter \(\alpha x_m\) and the input-state coefficients \(a\) and \(b\).
 Averaging uniformly over all possible target states on the Bloch sphere gives the average fidelity as~\cite{PhysRevA.69.022310}  (see Appendix B)
\begin{eqnarray}
\bar{\mathcal{F}}(x_m,\alpha)
&=&
\frac{3-e^{-4\sqrt{2}\alpha x_m}}{4}+
\frac{\left(1-e^{-4\sqrt{2}\alpha x_m}\right)^2}
{8e^{-2\sqrt{2}\alpha x_m}}
\nonumber\\&&\times\ln\!\left(
\frac{1+e^{-2\sqrt{2}\alpha x_m}}
{1-e^{-2\sqrt{2}\alpha x_m}}
\right).
\label{eq21}
\end{eqnarray}
\begin{figure}[t]
	\centering
	\includegraphics[width=0.46\textwidth]{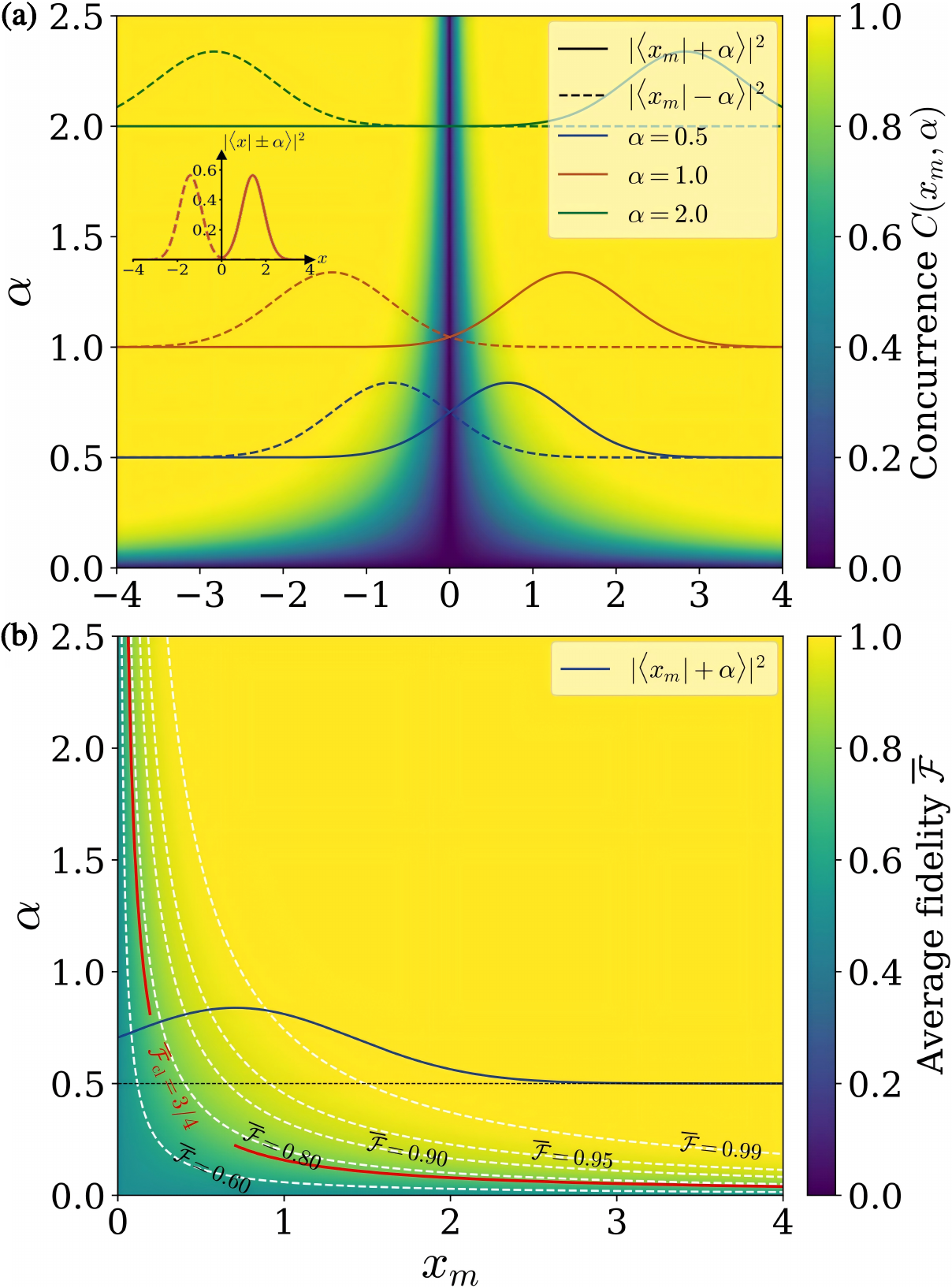}
	\caption{(a) Concurrence as a function of the coherent-state amplitude $\alpha$ and the measurement outcome $x_m$. The solid and dashed curves show the position-quadrature distributions of the coherent states $\lvert \alpha \rangle$ and $\lvert -\alpha \rangle$, respectively. The inset shows the amplitudes of the states $\ket{\pm\alpha}$ in the $x$ basis for $\alpha=1$. Different values of $\alpha$ yield the same amplitude profiles, differing only in the positions of the wave-packet centers. (b) Average fidelity of the RSP protocol as a function of the coherent-state amplitude $\lvert \alpha \rvert$ and the measurement outcome $x_m$. The solid red contour marks the classical benchmark of $3/4$, while the solid blue curve shows the position-quadrature distribution of the coherent state for $\alpha=0.5$.}
	\label{fig2}
\end{figure}
This result shows that the average fidelity is governed solely by the coherent-state amplitude \(\alpha\), through its effect on the measurement-outcome distribution. As \(\alpha\) increases,
the distribution of \(x_m\) is shifted toward larger values, thereby increasing the
typical value of \(\alpha x_m\) and thus the fidelity. Consequently, the fidelity
increases monotonically with \(\alpha\). In the large-amplitude limit, \(\alpha\gg 1\), the average fidelity approaches unity, $ \lim_{\alpha\to\infty}\bar{\mathcal{F}}(\alpha)\rightarrow 1$,
corresponding to the ideal remote-state-preparation limit. In the opposite limit, \(\alpha\to0\), we have $\lim_{\alpha\to 0}\bar{\mathcal{F}}(\alpha)=1/2$, since the relevant coherent states become indistinguishable owing to their complete wave-packet overlap. In Fig.~\ref{fig2}(b), we plot the mean fidelity as a function of the coherent-state amplitude $\alpha$ and the measurement outcome $x_m$. The results show that the fidelity can exceed the classical benchmark $\bar{\mathcal{F}}_{\mathrm{cl}}=3/4$ (see Appendix D) without requiring a large amplitude $\alpha$. Thus, high-fidelity state preparation remains attainable in the moderate-amplitude regime.

\section{Dissipation and physical implementation\label{sec4}}

\subsection{Effects of dissipation on the protocol}
So far, we have neglected imperfections in the system, which are inevitable in realistic implementations. We will consider two primary decoherence mechanisms: photon loss during fiber transmission and dephasing of the single spin.
The photon loss is modeled by a beam-splitter coupling with a vacuum environment, with transmittance $
\eta = \exp(-d_0/L_{\mathrm{att}}) $,
where $d_0$ is the transmission distance and $L_{\mathrm{att}}=25.5\,\mathrm{km}$ is the attenuation length of standard optical fiber at telecom wavelength \cite{PhysRevA.111.012438}.
The spin dephasing is described by a pure-dephasing rate $\gamma_2$, acting independently on each spin during the transmission time $\tau = d_0/c$ ($c\approx 2\times10^5\,\mathrm{km/s}$). Spin relaxation is neglected because the quantum information is typically encoded in long-lived ground-state spin levels and the relaxation time $T_1$ is assumed to be much longer than the timescale of the protocol.

We begin with the initial hybrid entangled state ${\rho}_{AB}(0)=|\Psi\rangle_{AB}\langle\Psi|$ of Eq. (\ref{eq6}). The two noise processes act on distinct physical degrees of freedom---the photonic mode and the spins, respectively---and are assumed to occur independently. Under the beam-splitter model, the signal mode $a_A$ couples to an environmental vacuum mode $b_E$, yielding the transformation $\ket{\pm\alpha}_A \ket{0}_E \to \ket{\pm\sqrt{\eta}\,\alpha}_A \ket{\pm\sqrt{1-\eta}\,\alpha}_E$.
Tracing over the environmental mode yields the spin-photon density matrix:
\begin{eqnarray}
\rho_{AB}^{\mathrm{refl}}
&=& \frac{1}{2}
\ket{\uparrow}_B\bra{\uparrow}
\otimes
\ket{\sqrt{\eta}\alpha}_A\bra{\sqrt{\eta}\alpha}
\nonumber\\
&&+ \frac{1}{2}
\ket{\downarrow}_B\bra{\downarrow}
\otimes
\ket{-\sqrt{\eta}\alpha}_A\bra{-\sqrt{\eta}\alpha}
\nonumber\\
&&+ \frac{1}{2}e^{-2(1-\eta)|\alpha|^2}
\nonumber\\
&&\times\Bigl(
\ket{\uparrow}_B\bra{\downarrow}
\otimes
\ket{\sqrt{\eta}\alpha}_A\bra{-\sqrt{\eta}\alpha}
+\mathrm{H.c.}
\Bigr).
\end{eqnarray}
Besides, the spin-dephasing channel acts on the off-diagonal terms as
\(\mathcal{E}_{\mathrm{spin}}\bigl(\ket{\uparrow}\bra{\downarrow}\bigr)=e^{-\gamma_2\tau}\ket{\uparrow}\bra{\downarrow}\) and
\(\mathcal{E}_{\mathrm{spin}}\bigl(\ket{\downarrow}\bra{\uparrow}\bigr)=e^{-\gamma_2\tau}\ket{\downarrow}\bra{\uparrow}\),
while leaving the diagonal terms unchanged.
Applying $\mathcal{E}_{\mathrm{spin}}\otimes \mathbb{I}_A$ to ${\rho}_{AB}^{\mathrm{refl}}$ gives the final spin-photon state after transmission:
\begin{eqnarray}
\rho_{AB}^{\mathrm{loss}}
&=&
\frac{1}{2}
\ket{\uparrow}_B\bra{\uparrow}
\otimes
\ket{\sqrt{\eta}\alpha}_A\bra{\sqrt{\eta}\alpha}
\nonumber\\
&&+ \frac{1}{2}
\ket{\downarrow}_B\bra{\downarrow}
\otimes
\ket{-\sqrt{\eta}\alpha}_A\bra{-\sqrt{\eta}\alpha}
\nonumber\\
&&+ \frac{1}{2}e^{-\gamma_2\tau}e^{-2(1-\eta)|\alpha|^2}
\nonumber\\
&&\times\Bigl(
\ket{\uparrow}_B\bra{\downarrow}
\otimes
\ket{\sqrt{\eta}\alpha}_A\bra{-\sqrt{\eta}\alpha}
+\mathrm{H.c.}
\Bigr).\label{eq23}
\end{eqnarray}
This result shows that the off-diagonal coherence is attenuated by two independent mechanisms. The factor
\(e^{-\gamma_2\tau}\) describes atomic dephasing, while
\(e^{-2(1-\eta)|\alpha|^2}\) results from photon loss. Indeed, the lost field carries the branch-dependent states
\(\ket{\pm\sqrt{1-\eta}\alpha}_E\), whose overlap is
${}_E\langle -\sqrt{1-\eta}\alpha | \sqrt{1-\eta}\alpha \rangle_E = e^{-2(1-\eta)|\alpha|^2}$.
Thus, increasing \(\lvert\alpha\rvert\) makes the environmental states more distinguishable and increases the which-branch information available to the environment, leading to stronger decoherence.
\begin{figure*}[t]
	\centering
	\includegraphics[width=0.98\textwidth]{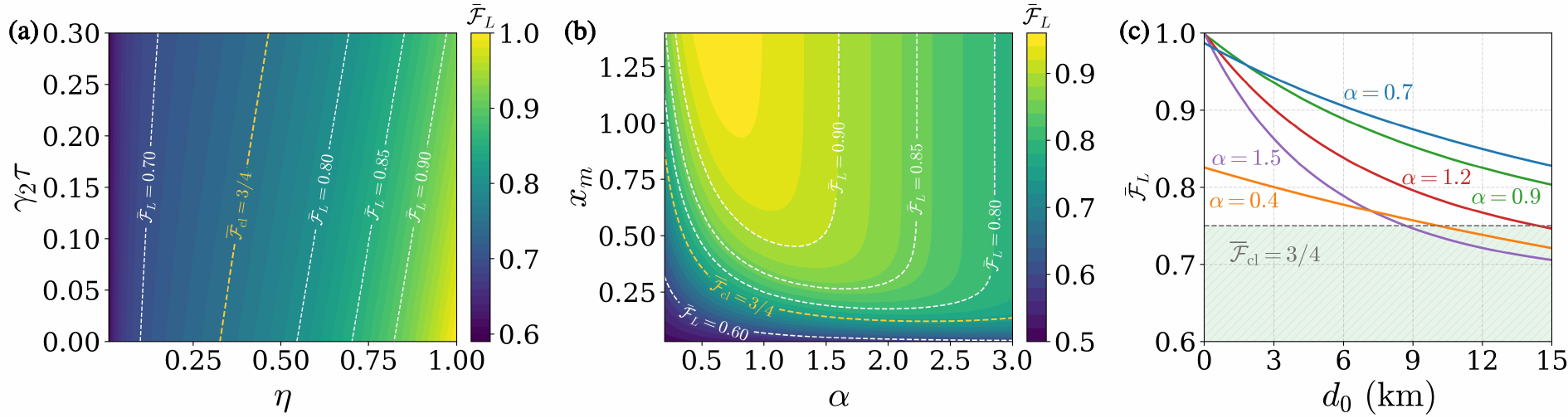}
	\caption{(a) Average fidelity of the RSP protocol as a function of the transmittance
 $\eta$ and the dephasing $\gamma_2\tau$ for $\alpha=1$ and $x_m=\sqrt{2}\alpha$. (b) Average fidelity of the RSP protocol as a function of the coherent-state amplitude $\alpha$ and the measurement outcome $x_m$ for $\eta=0.95$ and $\gamma_2\tau=0.1$. (c) Average fidelity as a function of the RSP distance $d_0$ for different values of the coherent amplitude $\alpha$. Here, we set the measurement result $x_m=\sqrt{2}\alpha$ and take the spin-dephasing rate to be $\gamma_2=2~\mathrm{kHz}$ and the optical attenuation length to be $L_{\mathrm{att}}=25.5~\mathrm{km}$.}
	\label{fig3}
\end{figure*}

Regarding the projection measurements, high-efficiency single-atom state
detection \cite{zhang2021high} and homodyne detection \cite{PhysRevLett.133.173604,xkt1-y58b} are experimentally accessible. We therefore assume, for simplicity, ideal projection measurements and neglect the effects of finite detection efficiency. This
simplification allows us to isolate and clearly identify the effects of
the dominant noise sources considered in our scheme. The effects of
imperfect measurements can be incorporated separately through an
appropriate detector-efficiency and measurement-error model~\cite{PhysRevA.111.012438}.
Following the ideal protocol, but with the noisy channel in Eq. \eqref{eq23}, the system state after the auxiliary atom interacting with the optical field is
\begin{eqnarray}
\rho_{ABA}^{\mathrm{loss}}
&=&
\frac{1}{4}
\ket{\uparrow}_{B}\bra{\uparrow}
\otimes
\Big[
\left(
\ket{\uparrow}_{A}\ket{\beta}_{A}
+
\ket{\downarrow}_{A}\ket{-\beta}_{A}
\right)
\nonumber\\
&&
\left(
\bra{\uparrow}_{A}\bra{\beta}_{A}
+
\bra{\downarrow}_{A}\bra{-\beta}_{A}
\right)\Big]
\nonumber\\
&&+
\frac{1}{4}
\ket{\downarrow}_{B}\bra{\downarrow}
\otimes
\Big[
\left(
\ket{\uparrow}_{A}\ket{-\beta}_{A}
+
\ket{\downarrow}_{A}\ket{\beta}_{A}
\right)
\nonumber\\
&&
\left(
\bra{\uparrow}_{A}\bra{-\beta}_{A}
+
\bra{\downarrow}_{A}\bra{\beta}_{A}
\right)\Big]
\nonumber\\
&&+
\frac{\Lambda}{4}
\ket{\uparrow}_{B}\bra{\downarrow}
\otimes
\Big[
\left(
\ket{\uparrow}_{A}\ket{\beta}_{A}
+
\ket{\downarrow}_{A}\ket{-\beta}_{A}
\right)
\nonumber\\
&&
\left(
\bra{\uparrow}_{A}\bra{-\beta}_{A}
+
\bra{\downarrow}_{A}\bra{\beta}_{A}
\right)
+\mathrm{H.c.}
\Big],
\label{eq24}
\end{eqnarray}
where we have defined the new parameters $
\beta=\sqrt{\eta}\,\alpha,
\Lambda=e^{-\gamma_{2}\tau}
e^{-2(1-\eta)|\alpha|^{2}}$. Homodyne detection then projects the state \eqref{eq24} into
\begin{eqnarray}
\widetilde{\rho}_{AB}
&=&\frac{1}{4}\Big[
 \ket{\uparrow}_{B}\bra{\uparrow}
 \otimes\ket{\chi_{+}}\bra{\chi_{+}}
+\ket{\downarrow}_{B}\bra{\downarrow}
 \otimes\ket{\chi_{-}}\bra{\chi_{-}}
\nonumber\\
&&+\Lambda\big(
 \ket{\uparrow}_{B}\bra{\downarrow}
 \otimes\ket{\chi_{+}}\bra{\chi_{-}}
+\mathrm{H.c.}\big)\Big],
\label{eq25}
\end{eqnarray}
where $\ket{\chi_{+}} = f_{+}\ket{\uparrow}_{A} + f_{-}\ket{\downarrow}_{A}$ and
$\ket{\chi_{-}} = f_{-}\ket{\uparrow}_{A} + f_{+}\ket{\downarrow}_{A}$,
with $f_{\pm}\equiv\braket{x_m}{\pm\beta}$. Finally, conditioned on obtaining the spin measurement outcome $\ket{{\mu}_{S0}}_A$, Bob's state is given by
\begin{eqnarray}
\widetilde{\rho}_{B}^{\rm{out}}
&=&\frac{1}{\mathcal{N}_{m}}
\Big[
|s_{+}|^{2}\ket{\uparrow}_{B}\bra{\uparrow}
+
|s_{-}|^{2}\ket{\downarrow}_{B}\bra{\downarrow}
\nonumber\\
&&+\Lambda s_{+}s_{-}^{*}
\ket{\uparrow}_{B}\bra{\downarrow}
+\mathrm{H.c.}
\Big]
\label{eq26}
\end{eqnarray}
with $s_{\pm}=af_{\pm}+bf_{\mp}$ and $\mathcal{N}_{m}
=|s_{+}(x_m,\alpha)|^{2}
+|s_{-}(x_m,\alpha)|^{2}$.
For $x_m \ge 0$, we further define
$\vartheta\equiv f_-/f_+ = \exp\big(-2\sqrt{2}\,\beta x_m\big).$
Corresponding to this state, the average RSP fidelity is given by (see Appendix C)
\begin{eqnarray}
\bar{\mathcal{F}}_{L}(x_m,\alpha)
&=&\frac{1}{{4\pi }}\int_0^\pi  {\sin\theta d\theta \int_0^{2\pi }}
{}_{B}\!\bra{\psi_{T}}
\widetilde{\rho}_{B}^{\rm{out}}
\ket{\psi_{T}}_{B}d\phi
\nonumber\\
 &=& \frac{{1 - \Lambda  + (8 + 4\Lambda ){\vartheta^2} - (1 + 3\Lambda ){\vartheta^4}}}{{16{\vartheta^2}}}\nonumber\\
 &&+ \frac{{{{(1 - {\vartheta^2})}^2}[(\Lambda  - 1) + (1 + 3\Lambda ){\vartheta^2}]}}{{32{\vartheta^3}}}\nonumber\\&&\times\ln \left( {\frac{{1 + \vartheta}}{{1 - \vartheta}}} \right),
\label{eq27}
\end{eqnarray}
Figure~\ref{fig3}(a) illustrates the effects of different noise sources on the performance of the protocol. It can be seen that fiber loss has a more pronounced effect than atomic dephasing. In Fig.~\ref{fig3}(b), we also show the protocol performance for different values of the coherent amplitude $\alpha$. For a given dephasing rate, the protocol exhibits an optimal value of $\alpha$. This behavior originates from the competition between the effects of the coherent amplitude in the absence and presence of channel loss. In the absence of fiber loss, increasing $\alpha$ generally improves the protocol performance, whereas in the presence of channel loss, smaller values of $\alpha$ are more favorable. As a result, an optimal coherent amplitude emerges when both atomic dephasing and fiber loss are taken into account.

This feature is shown more clearly in Fig.~\ref{fig3}(c), where the fidelity is plotted as a function of the preparation distance $d_0$ for different values of $\alpha$. Both excessively small and excessively large coherent amplitudes result in relatively poor fidelity, whereas an intermediate value of $\alpha$ leads to better performance. Notably, for a suitable range of coherent amplitudes, the protocol can still significantly exceed the classical fidelity limit of $3/4$ over relatively long preparation distances. In particular, for $\alpha=0.9$, the fidelity remains above $80\%$ when the preparation distance exceeds $15~\mathrm{km}$.

\subsection{Physical implementation}
The effective three-level configuration required by our protocol can, in
principle, be realized in various physical platforms, including trapped
neutral atoms~\cite{Saffman2010}, trapped ions~\cite{Blatt2008}, and
solid-state spin systems~\cite{Awschalom2013}. In the following, we focus
on a possible implementation based on a negatively charged nitrogen-vacancy
(NV) center in diamond. NV centers are attractive for this purpose because their electronic spins
provide long coherence times, high-fidelity microwave control, and optical
initialization and readout~\cite{doherty2013nitrogen,Jelezko2006,Balasubramanian2009}. Spin-photon entanglement
has also been demonstrated with NV centers, establishing a practical link
between stationary spin qubits and flying photons~\cite{Togan2010,Hensen2015}.

Let us consider the three-level manifold
\(\{\ket{\uparrow},\ket{\downarrow},|e\rangle\}\) of a negatively charged
NV center. The two lower states,
\(\ket{\uparrow}\equiv|m_s=-1\rangle\) and
\(\ket{\downarrow}\equiv|m_s=+1\rangle\), correspond to the Zeeman-resolved
components of the \(^{3}\!{A}_{2}\) ground-state triplet~\cite{PhysRevLett.113.020503,PhysRevLett.105.210501}. The upper state
\(|e\rangle\equiv\frac{1}{\sqrt{2}}(\ket{E_-}\ket{m_s=+1}+\ket{E_+}\ket{m_s=-1})\) is an optically excited state formed from
the spin states and orbital states $\ket{E_\pm}$~\cite{Ren17}. In the presence of a magnetic field, the two ground-state spin sublevels become spectrally resolved and are selectively coupled to the excited state by oppositely circularly polarized fields: the $|m_s=-1\rangle$ and $|m_s=+1\rangle$ states couple to $|A_2\rangle$ through the $\sigma^+$ and $\sigma^-$ transitions~\cite{Li20}, respectively. We therefore treat the NV center as an effective
\(\Lambda\)-type system in which the cavity mode selectively couples one
of the ground-state spin sublevels to the excited state \(|e\rangle\),
enabling the generation of the hybrid entangled state described by
Eq.~(\ref{eq6}). Meanwhile, a microwave field coherently couples the two
ground-state sublevels, allowing their quantum superpositions to be
prepared and manipulated arbitrarily.

For NV-center spins, the longitudinal relaxation time \(T_1\) is typically
much longer than the dephasing time \(T_2\)~\cite{doherty2013nitrogen}. At low temperatures, for
example, \(T_1\) can reach several minutes~\cite{PhysRevLett.110.156402}. In contrast, the dephasing
rate \(\gamma_2=1/T_2\) is approximately \(10~\mathrm{kHz}\) at room temperature~\cite{doherty2013nitrogen}
and can be reduced to the order of \(100~\mathrm{Hz}\) at low temperatures~\cite{PhysRevLett.110.156402}.
Consequently, as discussed above, the dominant limitation on RSP for NV spin is expected to be optical loss, rather than
intrinsic NV-spin relaxation or dephasing.
Moreover, recent experiments have demonstrated high-fidelity
single-shot readout of individual NV centers in diamond using
spin-to-charge conversion, with readout fidelities exceeding \(95\%\)~\cite{zhang2021high}. This capability provides a practical measurement interface
for the remotely prepared NV spin and makes high-fidelity state
verification feasible. Taken together, these results suggest that the proposed scheme could enable high-fidelity remote quantum-state preparation of NV centers within reach of current experimental capabilities.

\section{Conclusion\label{sec5}}
In summary, we have proposed and analyzed an RSP
protocol for a stationary electron spin based on hybrid entanglement between a
DV spin and a CV optical coherent state. The
hybrid resource is generated through spin-dependent reflection of a coherent
optical pulse from a spin-cavity system. At the receiver, the required projective
measurement on the optical subsystem is implemented by coupling the pulse to an
auxiliary spin, followed by homodyne detection and a target-state-dependent
projective measurement of the auxiliary spin. This approach avoids direct measurements of superpositions of nonorthogonal
coherent states, thereby substantially simplifying experimental implementation. For a
particular class of target states, the protocol enables deterministic RSP in the ideal limit, requiring only one bit of classical
communication for conditional state recovery.
A central feature of our scheme is the encoding of the quantum channel via spin-coherent-state entanglement. In contrast to spin-single-photon entanglement, where photon loss erases the transmitted photon and causes channel-distribution failure, a pure-loss channel merely attenuates the transmitted optical pulse.
Consequently, partial photon loss does not probabilistically erase the channel:
the transmitted pulse remains a well-defined coherent state and
available for the subsequent protocol.

We have also quantified the effects of coherent-state nonorthogonality, fiber loss, and spin dephasing on the preparation fidelity. Photon loss weakens the spin-optical entanglement and thus lowers the final RSP fidelity. Meanwhile, increasing the initial coherent-state amplitude improves the distinguishability of the optical branches while enhancing the which-branch information leaked to the environment, yielding an optimal amplitude for a given transmission distance. Under experimentally feasible parameters, the optimized average fidelity remains above the classical benchmark of
$3/4$ over tens of kilometers. These results, together with the feasibility of implementation in diamond nitrogen–vacancy centers, indicate that hybrid discrete-continuous-variable entanglement can combine the transmission advantages of optical fields with the storage and processing capabilities of stationary spins, offering a practical route toward resource-efficient RSP in distributed quantum networks.

	\begin{acknowledgments}
This work was supported by the Zhejiang Provincial Natural Science Foundation of China under Grant No. LMS25A040004 and the Innovation Program for Quantum Science and Technology (2023ZD0300904).
\end{acknowledgments}

\appendix
\setcounter{section}{0}
\renewcommand{\thesection}{\Alph{section}}
\setcounter{equation}{0}
\renewcommand{\theequation}{\thesection\arabic{equation}}

\stepcounter{section}
\section*{\textbf{APPENDIX A: CONCURRENCE OF THE CONDITIONAL TWO‑SPIN STATE}}
\addcontentsline{toc}{section}{APPENDIX A: CONCURRENCE OF THE CONDITIONAL TWO‑SPIN STATE}

In this appendix, we present a detailed derivation of the concurrence of the conditional
two‑spin state obtained after an $x$-quadrature homodyne
measurement.
After the optical pulse is measured and the homodyne outcome is obtained as \(x_m\), the two-spin system is projected onto the normalized conditional state [Eq. (\ref{eq14}) in the main text]
\begin{eqnarray}
{\left| {\Psi ({x_m})} \right\rangle _{AB}} &=& \frac{{{\psi _ + }({x_m})\left( {{{\left|  \uparrow  \right\rangle }_A}{{\left|  \uparrow  \right\rangle }_B} + {{\left|  \downarrow  \right\rangle }_A}{{\left|  \downarrow  \right\rangle }_B}} \right)}}{{\sqrt {2[|{\psi _ + }({x_m}){|^2} + |{\psi _ - }({x_m}){|^2}]} }}\nonumber\\
 &&+ \frac{{{\psi _ - }({x_m})\left( {{{\left|  \uparrow  \right\rangle }_A}{{\left|  \downarrow  \right\rangle }_B} + {{\left|  \downarrow  \right\rangle }_A}{{\left|  \uparrow  \right\rangle }_B}} \right)}}{{\sqrt {2[|{\psi _ + }({x_m}){|^2} + |{\psi _ - }({x_m}){|^2}]} }},\label{eqa1}
\end{eqnarray}
For a general pure two‑qubit state,
\begin{eqnarray}
\ket{\chi}
&=&
a_1\ket{\uparrow}_A\ket{\uparrow}_B
+
a_2\ket{\uparrow}_A\ket{\downarrow}_B
\nonumber\\&&+
a_3\ket{\downarrow}_A\ket{\uparrow}_B
+
a_4\ket{\downarrow}_A\ket{\downarrow}_B,
\end{eqnarray}
the concurrence is defined as
\begin{equation}
C(\ket{\chi})=2|a_1a_4-a_2a_3|,
\label{eqa3}
\end{equation}
which provides a direct measure of the entanglement between the two qubits. According to Eq. (\ref{eqa3}), the concurrence for the entangled state in Eq. (\ref{eqa1}) can be directly calculated to give
\begin{equation}
C(x_m,\alpha)
=
\frac{
	\left|
	\psi_{+}^{\,2}(x_m)-\psi_{-}^{\,2}(x_m)
	\right|
}{
	|\psi_{+}(x_m)|^2+|\psi_{-}(x_m)|^2
}.
\label{eq:general_conditional_concurrence}
\end{equation}
It is important to note that the numerator contains the squares, rather than
the modulus squares, of the probability amplitudes. In general,
\(\psi_{+}(x_m)\) and \(\psi_{-}(x_m)\) may possess a nontrivial relative phase,
so that \(\psi_{\pm}^{2}(x_m)\) and
\(\lvert\psi_{\pm}(x_m)\rvert^{2}\) are not equivalent. For real $\alpha$, however, $\psi_{\pm}(x_m)$ reduce to real-valued functions.
Then, the concurrence in Eq.~\eqref{eq:general_conditional_concurrence} can be simplified as
\begin{eqnarray}
C(x_m,\alpha)
=\left|
\frac{e^{4\sqrt{2}\alpha x_m}-1}{e^{4\sqrt{2}\alpha x_m}+1}
\right|
=\left| \tanh\left(2\sqrt{2}\alpha x_m\right) \right|.
\label{eqa5}
\end{eqnarray}
This is exactly the concurrence expression given in Eq.~\eqref{eq17} in the main text, which characterizes the entanglement available for the remote state preparation protocol.
Eq.~\eqref{eqa5} shows that the concurrence depends only on the product $\alpha x_m$ and satisfies the symmetry relations $
C(-x_m,\alpha)
=
C(x_m,-\alpha)
$. It vanishes along either $x_m=0$ or $\alpha=0$:
$
C(0,\alpha)=C(x_m,0)=0.
$

We now examine the physical meaning of these limits. At $x_m=0$, the two quadrature wave functions are equal, $\psi_+(0)=\psi_-(0)$, meaning that the two coherent-state components completely overlap in the position-quadrature space. Consequently, the homodyne measurement provides no which-path information about the photonic branch, and the conditional state becomes a product state
\begin{equation}
\begin{aligned}
|\Psi(0)\rangle_{AB}
&=\frac{|\Phi^+\rangle_{AB}+|\Psi^+\rangle_{AB}}{\sqrt{2}}
=|+\rangle_A|+\rangle_B,
\end{aligned}
\label{eq:x0_cond_state}
\end{equation}
where
\begin{equation}
\ket{+}
=
\frac{
	\ket{\uparrow}+\ket{\downarrow}
}{
	\sqrt{2}
}.
\end{equation}
Thus, the conditional state at $x_m=0$ is separable, and no entanglement is available for RSP.
In contrast, when $|\alpha x_m|\gg 1$, one of the two wave functions
dominates over the other, and
\begin{equation}
C(x_m,\alpha)\longrightarrow 1.
\end{equation}
In this regime, the two Gaussian peaks are well separated, and the homodyne outcome unambiguously identifies the photonic branch. The conditional spin state then approaches one of the maximally entangled Bell states, depending on the sign of the measurement result
\begin{equation}
\ket{\Psi(x_m)}_{AB}
\longrightarrow
\begin{cases}
\ket{\Phi^{+}}_{AB}, & x_m\gg 0,\\[4pt]
\ket{\Psi^{+}}_{AB}, & x_m\ll 0.
\end{cases}
\end{equation}
It should be emphasized that a large conditional concurrence does
not imply a large heralding probability. Although the
concurrence approaches unity for sufficiently large $|\alpha x_m|$,
measurement outcomes in the tails of the distribution $\lvert\psi_{\pm}(x_m)\rvert^2$ occur with low probability. Therefore, both the conditional concurrence and the homodyne probability should be taken into account when evaluating the practical performance of the entanglement-generation protocol.

\setcounter{equation}{0}
\stepcounter{section}
\section*{\textbf{APPENDIX B: Derivation of the average fidelity}}
\addcontentsline{toc}{section}{APPENDIX B: Derivation of the average fidelity}
\label{app:average_fidelity}

In this appendix, we evaluate the average fidelity of the proposed RSP protocol.
Given the optical homodyne outcome $x_m$ and the spin projective measurement $\ket{\mu_{S0}}_A$, the unnormalized conditional state of Bob is given by
\begin{eqnarray}
\bigl|\widetilde{\psi}_{B}^{\mathrm{out}}\bigr\rangle
&=&{}_A\!\langle\mu_{S0}|\Psi(x_m)\rangle_{AB}\nonumber\\
&=&\frac{\psi_{+}(x_m)a+\psi_{-}(x_m)b}{\sqrt{2\bigl[|\psi_{+}(x_m)|^2+|\psi_{-}(x_m)|^2\bigr]}} \ket{\uparrow}_B \nonumber\\
&&+\frac{\psi_{-}(x_m)a+\psi_{+}(x_m)b}{\sqrt{2\bigl[|\psi_{+}(x_m)|^2+|\psi_{-}(x_m)|^2\bigr]}} \ket{\downarrow}_B.
\label{app:unnormalized_bob_state}
\end{eqnarray}
For $x_m\geq 0$, it is convenient to introduce the ratio
\begin{equation}
\begin{aligned}
q\equiv \frac{\psi_{-}(x_m)}{\psi_{+}(x_m)}
=\exp\left(-2\sqrt{2}\alpha x_m\right),\qquad 0<q\le 1.
\end{aligned}
\label{eqb2}
\end{equation}
This ratio plays a central role in the subsequent fidelity analysis. It directly quantifies the overlap between the two coherent-state components in the position quadrature basis: for $q \ll 1$, the two Gaussian peaks are well separated and the coherent states are nearly distinguishable, whereas for $q \sim 1$, the peaks overlap significantly and the which-path information is largely erased. The restriction $x_m \ge 0$ is made without loss of generality, since the quadrature distributions are symmetric about $x_m = 0$, so that the analysis for negative $x_m$ is identical in form. Using the definition (\ref{eqb2}), Bob's normalized output state becomes
\begin{equation}
\begin{aligned}
\ket{\psi_B^{\mathrm{out}}}
=
\frac{
	\left(a+qb\right)\ket{\uparrow}_B
	+
	\left(qa+b\right)\ket{\downarrow}_B
}{
	\sqrt{
		1+q^2+2qs
	}
},
\end{aligned}
\label{app:normalized_bob_state}
\end{equation}
where
\begin{equation}
s
\equiv
ab^*+a^*b
=
2\operatorname{Re}(ab^*).
\label{eqb4}
\end{equation}
In the limit $q\rightarrow 0$, corresponding to well‑resolved
quadrature distributions, Eq.~\eqref{app:normalized_bob_state}
reduces to the target state
\begin{equation}
\ket{\psi_T}_B
=
a\ket{\uparrow}_B
+
b\ket{\downarrow}_B.
\label{app:target_state}
\end{equation}
For nonzero $q$, the prepared state will deviate from the target. To quantify the quality of the prepared state, we consider the following RSP fidelity:
\begin{equation}
\mathcal{F}
=
\bigl|_B\braket{\psi_T}{\psi_B^{\mathrm{out}}} \bigr|^2=
\frac{(1+qs)^2}{
	1+q^2+2qs
}.
\label{eqb6}
\end{equation}
Substituting Eq.~\eqref{eqb2} and Eq.~\eqref{eqb4} into Eq.~\eqref{eqb6}, we obtain
\begin{equation}
\begin{aligned}
\mathcal{F} = \frac{\Big(1 + 2\,\mathrm{Re}(ab^{*})\, e^{-2\sqrt{2}\alpha x_m}\Big)^2}
{1 + e^{-4\sqrt{2}\alpha x_m} + 4\,\mathrm{Re}(ab^{*})\, e^{-2\sqrt{2}\alpha x_m}}
\end{aligned}
\label{eq:20}
\end{equation}
To evaluate the fidelity averaged uniformly over the Bloch sphere, we
parameterize the input-state coefficients as
\begin{equation}
a
=
\cos\left(\frac{\theta}{2}\right),
\qquad
b
=
e^{i\phi}
\sin\left(\frac{\theta}{2}\right),
\label{app:bloch_parameterization}
\end{equation}
where $0\leq\theta\leq\pi$ and $0\leq\phi<2\pi$. With this parameterization, \(s=\sin\theta\cos\phi\), the fidelity averaged
uniformly over the Bloch sphere is given by
\begin{align}
\bar{\mathcal{F}}(q)
&=
\frac{1}{4\pi}
\int_{0}^{2\pi}d\phi
\int_{0}^{\pi}d\theta\,
\sin\theta\,
\mathcal{F}
\left(
q,s
\right).
\label{eqb9}
\end{align}
To evaluate this integral, we note that for a uniformly distributed Bloch vector, the quantity $s$ is itself uniformly distributed over the interval $[-1,1]$. This observation allows us to reduce the two-dimensional angular integral to a simpler one-dimensional integral over $s$, which greatly facilitates the analytical evaluation.
Hence, Eq.~\eqref{eqb9} can be reduced to the one‑dimensional integral
\begin{eqnarray}
\bar{\mathcal{F}}(q)
&=&
\frac{1}{2}
\int_{-1}^{1}ds\,
\frac{(1+qs)^2}{
	1+q^2+2qs
},\nonumber\\
&=&
\frac{3-q^2}{4}+\frac{(1-q^2)^2}{8q}\ln\left(\frac{1+q}{1-q}\right).
\label{eqb10}
\end{eqnarray}
Finally, using Eq. (\ref{eqb2}), the average fidelity can be
expressed directly in terms of $\alpha$ and $x_m$ as
\begin{equation}
\begin{aligned}
\bar{\mathcal{F}}(x_m,\alpha)
&=\frac{3-e^{-4\sqrt{2}\alpha x_m}}{4}\\
&\quad+\frac{\bigl(1-e^{-4\sqrt{2}\alpha x_m}\bigr)^2}{8e^{-2\sqrt{2}\alpha x_m}}
\ln\left[\frac{1+e^{-2\sqrt{2}\alpha x_m}}{1-e^{-2\sqrt{2}\alpha x_m}}\right].
\end{aligned}
\label{eqb12}
\end{equation}
This is exactly the average fidelity given in Eq.~\eqref{eq21} in the main text. It explicitly shows how the average fidelity depends on the measurement outcome and the coherent-state amplitude through the parameter $q$.
At $x_m=0$, one has $q=1$. Although the logarithmic term in Eq.~\eqref{eqb10} diverges as
$q\to 1^{-}$, its prefactor vanishes sufficiently rapidly that the product
remains finite. Therefore,
$\lim_{q\to 1^-}\bar{\mathcal{F}}(q)=1/2$.
Conversely, in the limit $\alpha x_m\to\infty$, for which $q\to 0^{+}$, one
finds $\lim_{q\to 0^+}\bar{\mathcal{F}}(q)=1$.
Thus, for $x_m>0$, the average fidelity increases from $1/2$ to unity as the
two quadrature distributions become increasingly distinguishable.

\setcounter{equation}{0}
\stepcounter{section}

\section*{\textbf{APPENDIX C: DERIVATION OF THE AVERAGE FIDELITY IN THE PRESENCE OF DECOHERENCE}}
\addcontentsline{toc}{section}{APPENDIX C: DERIVATION OF THE AVERAGE FIDELITY IN THE PRESENCE OF DECOHERENCE}
In this appendix, we extend the fidelity analysis of Appendix~B to account for
photon loss and spin dephasing.
For a given homodyne outcome $x_m$ and projective measurement $\ket{\mu_{S0}}_A$, Bob's conditional output state is described by the density matrix
\begin{eqnarray}
\widetilde{\rho}_{B}^{\rm{out}}
&=&\frac{1}{\mathcal{N}_{m}}
\Big[
|s_{+}|^{2}\ket{\uparrow}_{B}\bra{\uparrow}
+
|s_{-}|^{2}\ket{\downarrow}_{B}\bra{\downarrow}
\nonumber\\
&&+\Lambda s_{+}s_{-}^{*}
\ket{\uparrow}_{B}\bra{\downarrow}
+\mathrm{H.c.}
\Big],
\label{C1}
\end{eqnarray}
where we have defined the new parameters $s_{\pm}=af_{\pm}+bf_{\mp},\Lambda=e^{-\gamma_{2}\tau}
e^{-2(1-\eta)|\alpha|^{2}}$ and the normalization constant $\mathcal{N}_{m}
=|s_{+}(x_m,\alpha)|^{2}
+|s_{-}(x_m,\alpha)|^{2}$, with $f_{\pm}\equiv\braket{x_m}{\pm\sqrt{\eta}\alpha}$.
The fidelity $\mathcal{F}_L=_B\!\langle\psi_T|\rho_B^{\mathrm{out}}|\psi_T\rangle_B$ can be directly evaluated to give
\begin{equation}
\mathcal{F}_L
=
\frac{
	|a|^2|s_+|^2
	+
	|b|^2|s_-|^2
	+
	2\Lambda\,
	\mathrm{Re}
	\left(
	a^*b\,s_+s_-^*
	\right)
}{
	{\mathcal{N}_{m}}
}.
\label{C3}
\end{equation}
Parametrizing the target state on the Bloch sphere yields the average fidelity
\begin{eqnarray}
\mathcal{\bar{F}}_L(\theta,\phi)
&=&
\frac{N(\theta,\phi)}{\mathcal{N}_m},\\
\mathcal{N}_m
&=&
f_+^2+f_-^2+2f_+f_-\sin\theta\cos\phi,
\label{C4} \nonumber\\
N(\theta,\phi)
&=&
f_+^2
\left[
1+
\frac{\Lambda-1}{2}
\sin^2\theta
\right]
+
\frac{1}{2}
f_-^2\sin^2\theta
\nonumber\\
&&
+
(1+\Lambda)
f_+f_-\sin\theta\cos\phi
+
\frac{\Lambda}{2}
f_-^2\sin^2\theta\cos2\phi.\nonumber
\label{C5}
\end{eqnarray}
Considering the case $x_m>0$, we define the ratio
$\vartheta\equiv f_-/f_+$, satisfying $0\leq\vartheta<1$. Dividing both
$N$ and $\mathcal{N}_m$ by $f_+^2$ and introducing the variables
$s=\sin\theta\cos\phi$ and $v=\sin\theta\sin\phi$, for which
$\sin^2\theta=s^2+v^2$ and
$\sin^2\theta\cos 2\phi=s^2-v^2$, the average fidelity given by
Eq.~\eqref{eqb9} reduces to
\begin{equation}
\bar{\mathcal{F}}_L
=
\frac{1}{2}
\int_{-1}^{1}
ds\,
\frac{
	k_0+k_1s+k_2s^2
}{
	1+\vartheta^2+2\vartheta s
},
\label{C8}
\end{equation}
where
\begin{eqnarray}
k_0 &=& \frac{3+\Lambda+(1-\Lambda)\vartheta^2}{4}, \label{C9a} \nonumber\\
k_1 &=& (1+\Lambda)\vartheta, \label{C9b} \nonumber\\
k_2 &=& \frac{\Lambda-1+(1+3\Lambda)\vartheta^2}{4}. \label{C9c}\nonumber
\end{eqnarray}
Decomposing the integrand into partial fractions and carrying out the
integration, while noting that the odd term in $s$ vanishes, we obtain
\begin{equation}
\begin{aligned}
\bar{\mathcal{F}}_L
={}&
\frac{
	1-\Lambda
	+
	(8+4\Lambda)\vartheta^2
	-
	(1+3\Lambda)\vartheta^4
}{
	16\vartheta^2
}
\\
&+
\frac{
	(1-\vartheta^2)^2
	\left[
	(\Lambda-1)
	+
	(1+3\Lambda)\vartheta^2
	\right]
}{
	32\vartheta^3
}
\ln
\left(
\frac{1+\vartheta}
{1-\vartheta}
\right).
\end{aligned}
\label{C10}
\end{equation}
This is precisely the average fidelity given in Eq.~\eqref{eq27} in the main text.

\setcounter{equation}{0}
\stepcounter{section}
\section*{\textbf{APPENDIX D: THE CLASSICAL FIDELITY BENCHMARK FOR REMOTE STATE PREPARATION}}
\addcontentsline{toc}{section}{APPENDIX D: CLASSICAL BENCHMARK AVERAGE FIDELITY FOR REMOTE STATE PREPARATION WITH ONE CLASSICAL BIT}

In this appendix, we derive the optimal average fidelity achievable using a classical strategy, in which Alice has complete classical knowledge of the target pure-qubit state and may communicate at most one classical bit to Bob. A natural strategy is for Alice and Bob to agree in advance on two reference states that Bob can prepare. For each target state, Alice determines which of the two reference states has the higher fidelity with it and sends Bob one classical bit indicating her choice. Bob then prepares the corresponding reference state. We subsequently evaluate the average state-preparation fidelity achieved by this strategy.

The fidelity between a target pure state $\ket{\psi}$ and Bob's output
density matrix $\rho_B$ is defined as
\begin{equation}
\mathcal{F}(\psi,\rho_B)
=
\bra{\psi}\rho_B\ket{\psi}.
\label{D1}
\end{equation}
If Bob prepares a pure state $\ket{\varphi}$, this simplifies to
\begin{equation}
\mathcal{F}(\psi,\varphi)
=
\left| \braket{\varphi}{\psi} \right|^2.
\label{D2}
\end{equation}
Since the fidelity is linear in $\rho_B$, any randomization between
Bob's possible output states yields a fidelity equal to the
corresponding weighted average. This average cannot exceed the larger
of the individual fidelities. Therefore, in the present classical
strategy, Bob need only prepare one of the allowed pure reference
states; randomization cannot improve the fidelity. Therefore, the optimal output can always be taken as a pure state.
For an arbitrary pure state on the Bloch sphere as
\begin{equation}
\ket{\psi(\theta,\phi)}
=
\cos\frac{\theta}{2}\ket{0}
+
e^{i\phi}\sin\frac{\theta}{2}\ket{1},
\label{D3}
\end{equation}
with $0\le\theta\le\pi$ and $0\le\phi<2\pi$, the corresponding Bloch vector is $\bm r=(\sin\theta\cos\phi,\sin\theta\sin\phi,\cos\theta)$. The fidelity between two pure qubit states with Bloch vectors $\bm r$
and $\bm n$ is given by
\begin{equation}
\mathcal{F}(\bm r,\bm n)=\frac{1+\bm r\cdot\bm n}{2}.
\label{D4}
\end{equation}

Let us first consider preparing equatorial states of the form
\begin{equation}
\ket{\psi_\phi}
=
\frac{\ket{0}+e^{i\phi}\ket{1}}{\sqrt{2}},
\label{D5}
\end{equation}
with $\phi\in[0,2\pi)$, which can be prepared deterministically using the RSP protocol. Consider two antipodal reference states $\ket{+}_b$ and $\ket{-}_b$
on the equator. These two states divide the equatorial circle into two
semicircles. Alice sends one classical bit to tell Bob which reference
state is closer to the target state, and Bob prepares that state. Let $\delta$ denote the signed angular distance between the target state
and the closer reference state. Since the target is assigned to the
nearest reference state, $\delta\in[-\pi/2,\pi/2]$. The fidelity for a
given $\delta$ is
\begin{equation}
\mathcal{F}_{\max}(\delta)
=
\cos^2\frac{\delta}{2}
=
\frac{1+\cos\delta}{2}.
\end{equation}
Because the target states are uniformly distributed, $\delta$ is
uniformly distributed over an interval of length $\pi$. Therefore, the
average classical fidelity is
\begin{equation}
\bar {\mathcal{F}}_{\rm cl}^{\rm equator}
=
\frac{1}{\pi}
\int_{-\pi/2}^{\pi/2}
\cos^2\frac{\delta}{2}\,d\delta
=
\frac12+\frac1\pi\approx 0.8183.
\label{D6}
\end{equation}

For arbitrary pure qubit states uniformly distributed over the Bloch
sphere, the reference states correspond to the
north and south pole states. Alice uses a single bit to indicate
whether the target Bloch vector lies in the northern
($0\leq\theta\leq\pi/2$) or southern
($\pi/2\leq\theta\leq\pi$) hemisphere. Bob then prepares $\ket{0}$ or
$\ket{1}$ accordingly. The optimal fidelity for a given target state is
\begin{equation}
\mathcal{F}_{\max}(\theta)
=
\frac{1+|\cos\theta|}{2}.
\label{D8}
\end{equation}
The average fidelity over the Bloch sphere is therefore
\begin{eqnarray}
\bar {\mathcal{F}}_{\rm cl}^{\rm arbitrary}
&=&
\frac{1}{4\pi}
\int_0^{2\pi}d\phi
\int_0^\pi
\frac{1+|\cos\theta|}{2}\sin\theta\,d\theta \nonumber\\
&=&
\frac12
+
\frac14
\int_0^\pi
|\cos\theta|\sin\theta\,d\theta \nonumber\\
&=&
\frac34.
\label{D9}
\end{eqnarray}
We now show that
$\bar {\mathcal{F}}_{\rm cl}^{\rm arbitrary}=3/4$ is the global upper bound when
Bob's two output states are fixed pure states. Let $\bm n_0$ and
$\bm n_1$ denote their Bloch vectors. For a target Bloch vector $\bm r$,
Alice chooses the output state with the larger fidelity:
\begin{equation}
\mathcal{F}_{\max}(\bm r)
=
\max\left\{
\frac{1+\bm r\cdot\bm n_0}{2},
\frac{1+\bm r\cdot\bm n_1}{2}
\right\}.
\label{D10}
\end{equation}
Using $\max(x,y)=\tfrac12(x+y+|x-y|)$, we obtain
\begin{equation}
\mathcal{F}_{\max}(\bm r)
=
\frac12+\frac14\left[
\bm r\cdot(\bm n_0+\bm n_1)
+
\left|\bm r\cdot(\bm n_0-\bm n_1)\right|
\right].
\label{D11}
\end{equation}
Under the uniform Bloch-sphere average, $\langle\bm r\rangle=0$ and,
for any fixed vector $\bm u$,
$\langle|\bm r\cdot\bm u|\rangle=|\bm u|/2$. Hence,
\begin{equation}
\begin{aligned}
\bar {\mathcal{F}}_{\rm cl}
&=
\frac12
+\frac14
\left\langle
\left|\bm r\cdot(\bm n_0-\bm n_1)\right|
\right\rangle \nonumber\\
&=
\frac12+\frac{|\bm n_0-\bm n_1|}{8}.
\end{aligned}
\label{D12}
\end{equation}
Since $|\bm n_0|=|\bm n_1|=1$, we have
$|\bm n_0-\bm n_1|\leq 2$. Therefore,
\begin{equation}
\bar {\mathcal{F}}_{\rm cl}\leq
\frac12+\frac{2}{8}
=
\frac34.
\label{D13}
\end{equation}
Equality is achieved when $\bm n_1=-\bm n_0$, namely, when Bob's two
output states form an antipodal pair on the Bloch sphere.

	\bibliography{ref}
	
\end{document}